\documentclass[nofootinbib,twocolumn]{revtex4}
\usepackage{graphicx}
\usepackage{latexsym}
\def\be{\begin{equation}}
\def\ee{\end{equation}}
\def\bea{\begin{eqnarray}}
\def\eea{\end{eqnarray}}

\begin{document}
\title{A Single Scalar Field Model of Dark Energy with Equation of
State Crossing $-1$}
\author{Mingzhe Li}
\email{m.li@thphys.uni-heidelberg.de}
\affiliation{Institut f\"{u}r Theoretische Physik,
Philosophenweg 16, 69120 Heidelberg, Germany}
\author{Bo Feng}
\email{fengbo@mail.ihep.ac.cn}
\author{Xinmin Zhang}
\email{xmzhang@mail.ihep.ac.cn} \affiliation{Institute of High
Energy Physics, Chinese Academy of Sciences, P.O. Box 918-4, Beijing
100049, People's Republic of China}

\begin{abstract}
In this paper we study the possibility of building models of dark
energy with equation of state across $-1$ and propose explicitly a
model with a single scalar field which gives rise to an equation of
state larger than $-1$ in the past and less than $-1$ at the present
time, consistent with the current observations.
\end{abstract}

\maketitle

 \vskip 0.4cm

Since two groups \cite{Riess98,Perl99} independently showed the
accelerating expansion of our universe based on the Type Ia
Supernova (SN) observations of the redshift-distance relations in
1998, numerous observations and experiments consistently indicate
that our universe is dominated by a component with a negative
pressure, dubbed dark energy in the literature. Especially the first
year Wilkinson Microwave Anisotropy Probe (WMAP) data
\cite{Spergel03} strongly support for a concordance model in which
about $73\%$ of the energy of the universe is indwelt in the dark
energy ingredient. One essential to understand the nature of the
dark energy is to detect the evolution of its equation of state
$w=p/\rho$, where $p$ and $\rho$ are the pressure and the energy
density of the dark energy respectively. Recent analysis on the data
from the Supernova, cosmic microwave background (CMB) and large
scale structure (LSS) show that the cosmological constant fits well
to the data \cite{seljak}, however current data also mildly favor an
evolving dark energy with an equation of state $w$ larger than $-1$
in the past and less than $-1$ today \cite{starobinsky,ex} and evolving across
$-1$ in the intermediate redshift. If such a result holds on with
the accumulation of observational data, this would be a great
challenge to the current models of dark energy. Firstly, the
cosmological constant as a candidate for dark energy will be
excluded and dark energy must be dynamical. Secondly, the simple
dynamical dark energy models considered vastly in the literature
like the quintessence or the phantom can not be satisfied either.

In the quintessence models
\cite{wetterich,wetterich2,peebles,steinhardt}, $w$ is evolving
however always in the range of $-1 \leq w \leq 1$. In the phantom
models \cite{phantom}, one has $w \leq -1$. For the general
k-essence models $\mathcal{L}=p(\phi,~X)$ with $X=1/2
\nabla_{\mu}\phi\nabla^{\mu}\phi$, and $\nabla_{\mu}$ the covariant
derivative, it was shown that it also fails to cross the boundary of
$w=-1$ because of the problems of singularity and classical
instability in the perturbative equations \cite{zhao,vikman}. Thus
it is interesting to ask if theoretically there exists model of a
single scalar field which has a $w$ crossing $-1$ during the
evolution of the universe \footnote{The dark energy models with the
property of crossing the cosmological constant boundary are called
quintom models \cite{quintom}, and there are a lot of interests in
the phenomenological study of this class of models recently in the
literature
\cite{wei,vikman,zhang,guo,hu,xia,michael,li,cai,xfzhang}.}.
In this paper we propose a toy model of this type.

In the general k-essence models, the Lagrangian is a function of the
field $\phi$ and its first derivative $\nabla_{\mu}\phi$. In this
paper we will propose a dark energy model which includes higher
derivative operators in the Lagrangian, for example, \be
\mathcal{L}=\mathcal{L}(\phi,~X,~\Box\phi\Box\phi,~\nabla_{\mu}\nabla_{\nu}\phi\nabla^{\mu}\nabla^{\nu}\phi)~,
\ee where $\Box\equiv \nabla_{\mu}\nabla^{\mu}$ is the d'Alembertian
operator. The higher derivative terms in the effective Lagrangian
above may be derived from fundamental theories. In fact it has been
shown in the literature that this type of operators does appear as
the quantum corrections or due to the non-local physics in the
string theory \cite{simon,woodard,gross}. With the higher derivative
terms to the Einstein gravity, the theory is shown to be
renormalizable \cite{stelle} which has attracted many attentions. Recently, higher derivative operators 
have been considered to stabilize the linear fluctuations in the scenario of ``ghost condensation" \cite{ghost}. In
addition the canonical form for the higher derivative theory has
been put forward by Ostrogradski about one and a half century ago
\cite{ostrogradski}. In short, it is interesting and worthwhile to
study the implications of models with higher derivatives in
cosmology.

In this paper we will show this type of theories provides a class of
models of dark energy where the equation of state can get across
$-1$ during evolution, consistent with the current observations. We
consider the simple model with the Lagrangian given by \footnote{We
take the convention for the sign of the metric with $(+,~-,~-,~-)$
.} \be\label{lagrangian} \mathcal{L}=-{1\over
2}\nabla_{\mu}\phi\nabla^{\mu}\phi+{c\over
2M^2}\Box\phi\Box\phi-V(\phi)~, \ee where $M$ is a constant with
mass dimension and $c$ is a dimensionless constant. One can see that
this model differs from the phantom model \cite{phantom} due to the
dimension-6 operator. Our analysis below will show that this model
gives rise to an equation of state which transit from $w>-1$ to
$w<-1$ as the redshift decreases, furthermore it is equivalent to
the two scalar fields model of the quintom dark energy studied in
Ref.\cite{quintom,guo,xfzhang}.

Given the Lagrangian (\ref{lagrangian}) we have the
energy-momentum tensor:
\bea\label{stress} T^{\mu\nu}&=&[{1\over
2}\nabla_{\rho}\phi\nabla^{\rho}\phi+{c\over
2M^2}\Box\phi\Box\phi+{c\over
M^2}\nabla^{\rho}\phi\nabla_{\rho}(\Box\phi)\nonumber\\ & &+V]g^{\mu\nu}
-\nabla^{\mu}\phi\nabla^{\nu}\phi- {c\over
M^2}\nabla^{\nu}\phi\nabla^{\mu}(\Box\phi)\nonumber\\ & &- {c\over
M^2}\nabla^{\mu}\phi\nabla^{\nu}(\Box\phi), \eea and the equation
of motion: \be\label{eqm} -\Box\phi-{c\over
M^2}\Box^2\phi+\frac{dV}{d\phi}=0~. \ee
The energy-momentum tensor (\ref{stress}) can be rewritten as
 \bea\label{astress} T^{\mu\nu}&=&{1\over
2}[\nabla_{\rho}\psi\nabla^{\rho}\psi-\nabla_{\rho}\chi\nabla^{\rho}\chi
+2V(\psi-\chi)+\nonumber\\
& & {M^2\over c}\chi^2]g^{\mu\nu}
-\nabla^{\mu}\psi\nabla^{\nu}\psi+\nabla^{\mu}\chi\nabla^{\nu}\chi~,
\eea where $\chi$ and $\psi$ are defined by
 \bea
\chi &=&\frac{c}{M^2}\Box\phi\label{change}~,\\
\psi &=&\phi+\chi~. \eea
It is not difficult to see that
the energy-momentum tensor
(\ref{astress}) can be derived from the following Lagrangian
\be\label{alagrangian} \mathcal{L}= -{1\over
2}\nabla_{\mu}\psi\nabla^{\mu}\psi+{1\over
2}\nabla_{\mu}\chi\nabla^{\mu}\chi -V(\psi-\chi)-{M^2\over
2c}\chi^2~, \ee
with $\psi$ and $\chi$ being two independent
fields. The variations of the Lagrangian in (\ref{alagrangian}) with
respect to the fields $\psi$
and $\chi$ respectively give rise to the following equations of motions,
\bea
& &\Box\psi-V'=0~,\\
& &\Box\chi+\frac{M^2}{c}\chi-V'=0~, \eea where the prime denotes the
derivative with respect to $\psi-\chi$.

When we take $\psi =\phi+\chi$ in (8) and (9),
we recover the equation of motion (\ref{eqm}) of the single field
model and the transformation equation (\ref{change}). So, one can
see from the derivations above that our single field model
(\ref{lagrangian}) is equivalent to the one with two fields (\ref{alagrangian}) where one is the
normal (quintessence-like) scalar field $\chi$ and another is the ghost
(phantom-like) field $\psi$. 
In general, the normal mode $\chi$ gives rise to the equation of
state $w_1 \geq -1$ and the ghost mode has $w_2 \leq -1$,
consequently the energy density in $\chi$ would decrease and that in
$\psi$ would increase with the expansion of the universe. Assuming
$\chi$ dominates the dark energy sector in the early time, the
equation of state for the system will be larger than $-1$ during
that period of time. When the ghost mode becomes dominant over the
normal one (and will continue to be dominant among all the
components in the universe), the equation of state of the system
will become $<-1$ and cross the cosmological constant boundary at a
intermediate redshift. Quantitatively the moment when the equation
of state crosses $-1$ depends on the values of the model parameters.

Now we study the perturbations of our model.
As is well known, in the usual quintessence and phantom models,
the field and metric perturbations are stable
on the small length scales. For both the quintessence and the phantom
models the sound speed as indicated in Ref. \cite{kinflation} is given by
\be
c_s^2= \frac{\partial p/ \partial X}{\partial \rho/ \partial X}=1~,
\ee
where
$X\equiv (1/2)\nabla_{\mu}\phi\nabla^{\mu}\phi$.
This can be seen clearly from the dispersion relations. For the phantom
field, the frequency $\omega$ relates to the momentum $\textbf{k}$
in the same way as in the case of the quintessence,
$\omega^2=\textbf{k}^2+...$. And the sound speed is often defined as
$\omega^2=c_s^2\textbf{k}^2+...$.
Since we have demonstrated that our model is equivalent to a model with
a quintessence-like and a phantom-like scalar fields, the general
arguments for the quintessence and the phantom hold here, which indicate
 there will be no exponentially
instabilities of the perturbations on small scales in our model.  

In general the potential term $V(\psi-\chi)$ (equivalently $V(\phi)$
in our single field model) should include the interactions between
the two fields $\psi$ and $\chi$. For some specific choices of the
potential $V$, however these two modes could decouple. As an example
we consider $V=(1/2)m^2\phi^2$. Then the Lagrangian
(\ref{alagrangian}) can be ``diagonalized" as
\be\label{aalagrangian} \mathcal{L}={1\over
2}\nabla_{\mu}\phi_1\nabla^{\mu}\phi_1 -{1\over
2}\nabla_{\mu}\phi_2\nabla^{\mu}\phi_2-{1\over 2}m_1^2\phi_1^2
-{1\over 2}m_2^2\phi_2^2~, \ee through the transformation \be \left(
\begin{array}{c} \psi \\ \chi\end{array} \right)
=\left(\begin{array} {cc} -a_1 & a_2\\-a_2 & a_1\end{array}\right)
\left( \begin{array}{c} \phi_1\\ \phi_2
\end{array} \right)~.
\ee In (12) and (13) \bea & & a_1={1\over 2}(1+\frac{4c
m^2}{M^2})^{-1/4}
(\sqrt{1+\frac{4c m^2}{M^2}}-1)~,\nonumber\\
& & a_2={1\over 2}(1+\frac{4c m^2}{M^2})^{-1/4}(\sqrt{1+\frac{4c m^2}{M^2}}+1)~,
\eea
and
\bea\label{mass}
& &m_1^2=\frac{M^2}{2c}(\sqrt{1+\frac{4cm^2}{M^2}}+ 1)~,\nonumber\\
& &m_2^2=\frac{M^2}{2c}(\sqrt{1+\frac{4cm^2}{M^2}}- 1)~.
\eea
Hence, one can see that our model
 (\ref{lagrangian}) with $V=(1/2)m^2\phi^2$ is
equivalent to the uncoupled system
(\ref{aalagrangian}). The two modes
$\phi_1$ and $\phi_2$ evolve independently in the
universe. The positivity of the parameter $c$
guarantees the positivity of $m^2_1$ ($m^2_2$ is always positive
as long as $1+\frac{4c m^2}{M^2}>0$). In the limit of $m\ll M$, the
masses of the two modes are approximately $m_1\simeq M/\sqrt{c}$ and
$m_2\simeq m$.
In fact, $\phi_1$ and $\phi_2$ are the eigenfunctions of the
d'Alembertian operator, $\Box$, with the
eigenvalues $-m_1^2$ and $m_2^2$ respectively. The solution to the equation of motion
(\ref{eqm}) $\phi$ is decomposed by these eigenfunctions as
$\phi=(1+\frac{4c m^2}{M^2})^{-1/4}(\phi_1+\phi_2)$.

As we mensioned above, the perturbations in the phantom component $\phi_2$ present no unphysical instabilities in the classical level. 
In fact, the spatial fluctuation of $\phi_2$ with wavenumber $k>m_2$ is stable, the solution to the equation of motion
is oscillatory with time. The time-rising behavior only presents on very large scales, $L>1/m_2$. If the phantom mass $m_2$ of the phantom field is smaller than $H$, there will be no instability within the horizon \cite{hsu}. 
The rising behaviors of the super-horizon modes (including the background mode, which corresponds to $k=0$) of phantom do not matter in cosmology. This is because the universe is expanding instead of static, generally the components in the universe are evolving. Furthermore, the Hubble expansion provides a friction force which prevents these modes from increasing exponentially. In another word, such an instability is ``benign" \cite{smilga}.

\begin{figure}[htbp]
\includegraphics[scale=0.85]{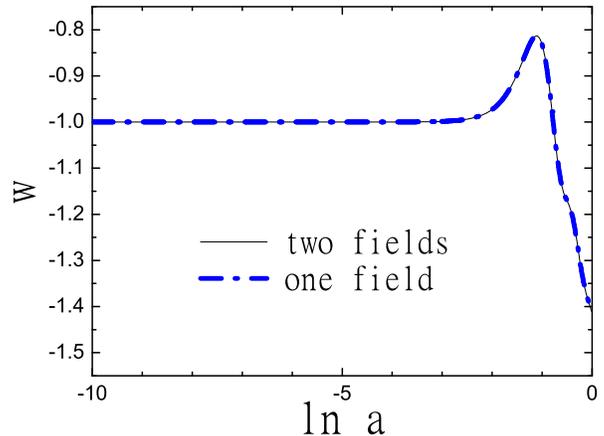}
\caption{\label{equivalence}Plots of the equation of state
for our model and the two-field model with one
quintessence and one phantom field. In the numerical calculations we
choose the initial conditions given
by Eq. (\ref{equiv}) and have set $m=2\times 10^{-61} M_{pl}$,
$\frac{c}{M^2}=10^{120} M_{pl}^{-2}$. The equivalent initial
conditions for the two scalar fields are $\phi_1(0)=3\times
10^{-2}M_{pl}$, $\phi_2(0)=0.18M_{pl}$, and
$\dot{\phi_1}(0)=0$, $\dot{\phi_2}(0)=0$.
$\Omega_{DE}\approx 0.73$ and $h\approx 0.68$. }.
\end{figure}

 In Fig.\ref{equivalence} we plot the equation of state of our model as a function of $\ln a$.
One can see clearly $w$ crosses $-1$ during evolution. Furthermore,
we made a test on the equivalence of the model (\ref{lagrangian})
with the two-field quintom model numerically for $V=(1/2)m^2\phi^2$.
If the background spacetime is described by the
Friedmann-Robertson-Walker metric,
$ds^2=dt^2-a^2(t)\delta_{ij}dx^idx^j$ and the field $\phi$ is
homogeneous, the d'Alembertian operator will depend only on the time
derivatives, $\Box=\partial^2/\partial t^2+3H\partial/\partial t$.
The equation of motion (\ref{eqm}) becomes: \be
-\ddot\phi-3H\dot\phi+m^2\phi-{c\over
M^2}(\frac{\partial^2}{\partial t^2}+3H\frac{\partial}{\partial
t})(\ddot\phi+3H\dot\phi)=0~, \ee where dot represents the
derivative with respect to time. The energy density and the pressure
are \be\label{energy} \rho=-{1\over 2}\dot\phi^2+{c\over
2M^2}(\ddot\phi+3H\dot\phi)^2-{c\over M^2}\dot\phi\frac{\partial}
{\partial t}(\ddot \phi+3H\dot \phi)+{1\over 2}m^2\phi^2~; \ee
\be\label{pressure} p=-{1\over 2}\dot\phi^2-{c\over
2M^2}(\ddot\phi+3H\dot\phi)^2-{c\over
M^2}\dot\phi\frac{\partial}{\partial t} (\ddot \phi+3H\dot
\phi)-{1\over 2}m^2\phi^2~, \ee respectively.
The initial conditions picked up in the numerically calculations are
\bea\label{equiv}
\phi(0)&=& g[\phi_1(0)+\phi_2(0)]~,\nonumber\\
\dot\phi(0)&=& g[\dot{\phi}_1(0)+\dot{\phi_2}(0)]~,\nonumber\\
\ddot \phi(0)&=& g[-3H(0)(\dot{\phi}_1(0)+\dot{\phi}_2(0))-m_1^2\phi_1(0)+
m_2^2\phi_2(0)],\nonumber\\
\dddot \phi(0)&=& g[(9H(0)^2-3\dot H(0))(\dot{\phi}_1(0)+\dot{\phi}_2(0))+\nonumber\\
&&3H(0)(m_1^2\phi_1(0)-m_2^2\phi_2(0))-\nonumber\\
& &(m_1^2\dot{\phi}_1(0)-m_2^2\dot{\phi}_2(0))]~,\nonumber\\
H^2(0)&=& \frac{8\pi G}{3}[\rho_m(0)+\rho_r(0)+{1\over 2}\dot{\phi_1^2}(0)-{1\over 2}
\dot{\phi_2^2}(0)
+\nonumber\\ & &{1\over 2}m_1^2\phi_1^2(0)+
{1\over 2}m_2^2\phi_2^2(0)]~,\nonumber\\
\dot H(0)&=&-4\pi G
[\rho_m(0)+\frac{4}{3}\rho_r(0)+\dot{\phi_1^2}(0)-\dot{\phi_2^2}(0)]~,
\eea where $0$ in the bracket means the initial time, $\rho_m$ and
$\rho_r$ are the energy densities of matter and radiation
respectively, and $g$ is defined as \be g\equiv (1+\frac{4c
m^2}{M^2})^{-1/4}~. \ee We depicte the numerical result for the
comparison in Fig.\ref{equivalence}. One can see from this figure
the equation of state from (\ref{energy}) and (\ref{pressure}) of
our single scalar field model does coincide with that of the
two-field model (\ref{aalagrangian}).

In summary, we in this paper have proposed a single scalar field
model of dark energy which has the property of crossing the
cosmological constant boundary. This is achieved by considering
higher derivative operators. And this class of model is free from
the difficulties of the singularity and the gravitational
instabilities in the general k-essence-like models. We should point
out that although the energy-momentum tensor of this single scalar
field model is equivalent to two independent scalar
fields\footnote{After our paper appeared, the authors of
Ref.\cite{binf} used the "$\Box$" term only and dropped the
conventional kinetic term as well the potential, which cannot be
identified with two scalar fields. }, when we consider the
interactions they may possibly show some different behaviors. It
might be theoretically possible to solve the inherent problems, the
quantum instabilities exhibited in the model of
phantom\cite{Phtproblms} and is worth studying further.

Our study here has the following implications: 1) in the general
phenomenological fitting appearing in the literature of
the dark energy models
to the observational data such as SN, CMB, LSS an issue is how
to consistently include the perturbations of the dark energy. The key to
this issue is to build theoretical models of dark energy with the equation
of state crossing $-1$. Our model proposed in this paper serves as an
example for this class of
models of dark energy to study the cosmological implications numerically;
2) similar to the k-inflation our model can be
applied for inflation. And in this case one would expect non-vanishing isocurvature
perturbations even though this model includes explicitly only one inflaton
field.

{\bf Acknowledgements:} M.L. would like to thank Michael Doran and
Christof Wetterich for discussions and to acknowledge the support
from the Alexander von Humboldt Foundation. B.F. and X.Z. would like
to thank Xiao-Jun Bi, Hong Li and Yun-Song Piao for discussions and
are supported in part by the National Natural Science Foundation of
China under the grant No. 90303004 and also by the Ministry of
Science and Technology of China under grant No. NKBRSF G19990754.

\end{document}